\documentclass[twocolumn,amsmath,amssymb,floatfix]{revtex4}

\usepackage{bm}
\usepackage{amsmath}
\usepackage{epsf}

\newcommand{\beq}{\begin{equation}}

\newcommand{\eeq}{\end{equation}}
\newcommand{\beqa}{\begin{eqnarray}}
\newcommand{\eeqa}{\end{eqnarray}}

\newcommand{\bk}{{\bf k}} 
\newcommand{\bx}{{\bf x}}

\newcommand{\ApJS}{Astrophys. J Supp.}
\newcommand{\ApJL}{Astrophys. J Lett.}
\newcommand{\ApJ}{Astrophys. J}

\newcommand{\PRD}{Phys. Rev. D}
\newcommand{\MNRAS}{Mon. Not. R. Astron. Soc.}

\newcommand{\AsAs}{Astron. Astrophys.}

\newcommand{\etal}{{\it et al. }}
\newcommand{\aut}[2]{{#2.\ #1,}}
\newcommand{\paut}[2]{{#2.\ #1} and}
\newcommand{\laut}[2]{{#2.\ #1,}}
\newcommand{\refs}[6]{#2, {\bf #3},  {#4} (#5).}
\newcommand{\mrefs}[6]{#2, {\bf #3},  {#4} (#5);}

\newcommand{\mybib}[2]{\bibitem{#2}}

\begin{document}

\title{Self-Calibration of Cluster Dark Energy Studies: Counts in Cells}
\author{Marcos Lima$^{1}$ and Wayne Hu$^{2}$}
\affiliation{
$^{1}$Department of Physics, University of Chicago, Chicago IL 60637\\
$^{2}$Center for Cosmological Physics, Department of Astronomy and Astrophysics,
and Enrico Fermi Institute, University of Chicago, Chicago IL 60637 \\}

\begin{abstract}
\baselineskip 11pt
Cluster number counts can constrain the properties of dark energy if and only if 
the evolution in the relationship between observable
quantities and the cluster mass can be calibrated.
Next generation surveys with $\sim 10^{4}$ clusters
will have sufficient statistics to enable
some degree of self-calibration. The excess variance of counts 
 due to the clustering of clusters provides such an opportunity
 and can be measured from the survey without additional observational cost.  It can
minimize the degradation in dark energy constraints due to an unknown 
power law evolution in the mass-observable relation improving 
constraints on the dark energy equation of state by a factor of 2 or more to
$\sigma(w)=0.06$ for
a deep 4000 deg$^{2}$ survey.
\end{abstract}
\maketitle

\section{Introduction}

The abundance of clusters of galaxies of a known mass
is a sensitive probe of linear density fluctuations 
given Gaussian random initial
conditions (e.g. \cite{GenRef}).
Since the growth of fluctuations is halted by the dark energy, abundance measurements
as a function of redshift 
can be translated into dark energy constraints \cite{WanSte98,HaiMohHol01}.  

Unfortunately,
the mass of a cluster is not a direct observable.
The key to exploiting this dark energy sensitivity
is to determine the relationships between observable
quantities such as flux or temperature and the mass.
{\it Ab initio} computations from cosmological simulations serve as useful guides
 but invite misinterpretation when used directly 
due to missing gas, star formation, and AGN physics (e.g. \cite{PieScoWhi01,Sel02}).
Much like with the distance ladder determinations of the Hubble constant, more progress
can be made by cross-calibrating the mass-observable relations between 
$X$-ray, lensing, optical and microwave
surveys.   A potential drawback
is that cross-calibration makes the interpretation of each data set subject to the systematic
errors in all of the surveys employed and so the weakest link in the chain.   
Cross-calibration can also only be performed across the redshift and mass range where surveys
overlap.

Recently, 
self-calibration techniques have been advocated as a useful check on cross-calibration 
and the internal consistency of dark energy determinations.  
Clusters benefit from having several mass-sensitive properties that can be 
reliably predicted from simulations of their dark matter properties only.
The first of course is the evolution of the abundance itself.  
An unknown constant normalization of the mass-observable relation
can be determined from the evolution in the abundance above threshold
in the given observable \cite{LevSchWhi02}.
Even an arbitrarily evolving normalization can be determined by measuring the abundance
as a function of the threshold and redshift \cite{Hu03a}.  
This technique however requires a dynamic range in the mass function that is substantially larger than
the scatter in the mass-observable relation and will only be possible with very deep surveys or
at fairly low redshifts.
Finally, mild evolution in the mass-observable relation can be
calibrated by examining the power spectrum of the clusters \cite{MajMoh03}.  
The drawback of the latter method is that
measuring the three dimensional power spectrum generally requires precise redshifts and hence
costly spectroscopy or an improvement in photometric redshift techniques beyond that required
for abundance studies themselves $(\Delta z \approx 0.1)$.

Here we point out that there is an analog of the power spectrum self-calibration
technique that will automatically come out of the statistical analysis of the number counts in
any given cluster survey.  The excess or sample 
variance of the counts  due 
to the clustering of clusters must be included in future error analyses \cite{HuKra02}.  
For self-calibration
purposes, this source of  ``noise''  is actually signal and can be used to calibrate
the mass observable relation.

\section{Noise as Signal}

The probability of measuring a 
number of clusters $N$ in a cell of a given redshift $z$ and angular extent
is given by the Poisson distribution
\begin{equation}
P(N | m) = {m^N \over N!} e^{-m}\,,
\end{equation}
where the mean $m \equiv \langle N \rangle_P$ and the brackets denote averaging over realizations
of the Poisson process.

Now take a set of cells indexed by $i$ where
 the mean number $m_i$ fluctuates in space
\begin{equation}
m_i = \int d^3 x W_i(\bx) n({\bx};z_i)\,,
\end{equation}
where  $W_i(\bx)$ is the cell window function, in this case a top hat,  and $n(\bx;z_i)$ is the
spatial number density.   On large scales, 
fluctuations in their spatial number density trace the
linear density fluctuations $\delta(\bx;z)$ from the
large scale structure of the universe
\begin{equation}
n(\bx;z) = \bar n(z) [1+ b(z) \delta(\bx;z)]\,,
\label{eqn:nz}
\end{equation}
where $b$ is the linear bias of the clusters.  
Overbars denote a spatial average or a sample average over realizations of the large 
scale structure.   Thus the sample averaged number counts
$\bar m_i = V_i \bar n(z_i)$ where the cell volume is $V_i \equiv \int d^3 x W_i$.

The mean numbers then possess a sample covariance given by the linear power spectrum
$P(k)$ \cite{HuKra02}
\begin{eqnarray}
S_{ij} &=&\langle (m_i -\bar m_i)(m_j - \bar m_j)\rangle_S \nonumber\\
&=&{ b_i \bar m_i b_j \bar m_j  \over V_i V_j} \int{d^3 k \over (2\pi)^3} W_i^*(\bk)W_j(\bk) P(k)\,.
\label{eqn:covariance}
\end{eqnarray}
Here $W_i(\bk)$ is the Fourier transform of $W_i(\bx)$ 
and $b_i = b(z_i)$.   For a single spherical cell of comoving radius $R$,
the fractional errors 
$S_{ii}^{1/2}/m_i = b_i \sigma_R$ where $\sigma_R$ is the rms linear density fluctuation
in the cell.

The likelihood of drawing a set of cluster counts ${\bf N}=(N_1,...N_c)$
given a model for  $\bar{\bf m}$ and ${\bf S}$ is then
\begin{eqnarray}
L({\bf N}| \bar {\bf m}, {\bf S}) = \int d^c m  \left[\prod_{i=1}^c 
P(N_i | m_i) \right] G({\bf m} | \bar{\bf m}, {\bf S})\,,
\label{eqn:likelihood}
\end{eqnarray}
where $G$ denotes the multivariate Gaussian distribution
\begin{eqnarray}
G({\bf m} | \bar{\bf m}, {\bf S}) = {1 \over \sqrt{(2\pi)^{c} {\rm det} {\bf S}} }e^{-{1\over 2} ({\bf m} - \bar{\bf m}) {\bf S}^{-1} ({\bf m} - \bar{\bf m})}\,.
\end{eqnarray}

It is instructive to consider a few special cases.  
In the limit that sample variance is negligible compared with the
 Poisson variance $S_{ii} \ll \bar m_i$
the Gaussian approximates a delta function and
\begin{eqnarray}
L({\bf N}| \bar {\bf m}, {\bf S})\approx \prod_{i=1}^c
P(N_i| \bar m_i)\,.
\end{eqnarray}
This form is appropriate for very rare clusters and is used in the analysis
of local high temperature clusters.

In the limit of large numbers $m_i \gg 1$, the Poisson distribution approaches 
a Gaussian 
\begin{equation}
\prod_{i=1}^c  P(N_i | \bar m_i ) \approx  G({\bf N} | {\bf m},{\bf M})\,,
\end{equation}
where ${\bf M} = {\rm diag}({\bf m}_i)$.
The likelihood becomes a convolution of Gaussians
\begin{eqnarray}
L({\bf N}| \bar {\bf m}, {\bf S}) \approx \int d^c m \, G({\bf N} | {\bf m},{\bf M}) G({\bf m} | \bar{\bf m}, {\bf S}) \,,
\end{eqnarray}
or via the convolution theorem 
approximately a Gaussian in ${\bf N}-\bar{\bf m}$ with covariance ${\bf C} \equiv
{\bf S}+\bar {\bf M}$.

The statistical properties of the counts are hence specified by their mean
 $\bar{\bf m}$ and their sample covariance ${\bf S}$.  The space density $\bar n(z)$ 
controls $\bar{\bf m}$ and hence the ``signal"; the bias $b(z)$ controls 
${\bf S}$ and hence the ``noise".  Given a cosmology, both 
$\bar n$ and $b$ can be predicted as a function of the cluster
mass (see \S \ref{sec:self}).  From the perspective of mass calibration both
ingredients are therefore signal.

\begin{figure}[tb]
\centerline{\epsfxsize=3.2in\epsffile{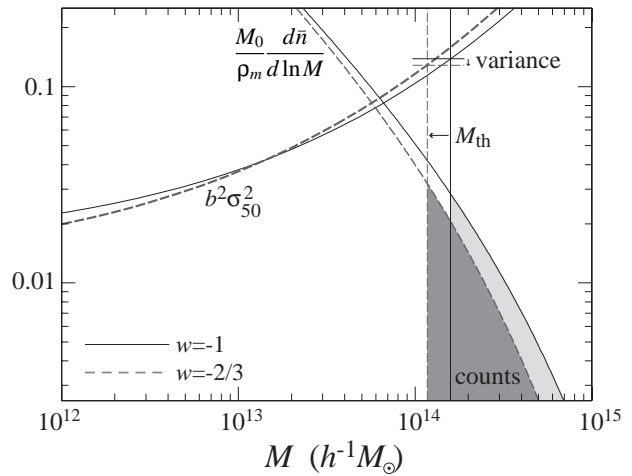}}
\caption{\footnotesize
Mass sensitivity of the counts and their variance compared with
sensitivity to the dark energy equation of state $w$ at $z=0$.  The integral of the mass
function $d\bar n/d\ln M$, here normalized to $M_0=10^{14} h^{-1} M_\odot$ from the
 threshold
$M_{\rm th}$ predicts
the expected counts.  Changes due to $w$ can be
compensated by a comparable shift in $M_{\rm th}$ (dark vs.\ light shaded region) but also
change the fractional cell variance at threshold (horizontal lines), shown here normalized to the typical
cell volume ($50 h^{-1}$ Mpc radius)  as $b^2 \sigma_{50}^2$.}
\label{fig:massfn}
\end{figure}

\section{Fisher Matrix}

The Fisher matrix formalism provides a means of making projections for
the result of likelihood analyses of future data given a parameterized model for the
mean $\bar{\bf m}$ and sample covariance ${\bf S}$ of the counts as well as
a fiducial choice of the true parameters.
 
Given a set of parameters $p_\alpha$ on which these quantities depend,
the information is quantified by the Fisher matrix
\begin{equation}
F_{\alpha\beta} \equiv -\langle { \partial^2 \ln L \over \partial p_\alpha \partial p_\beta} 
\rangle
\end{equation} 
such that the parameter covariance matrix  $C_{\alpha\beta} \approx ({\bf F}^{-1})_{\alpha\beta}$.
The marginalized parameter errors are then $\sigma(p_\alpha) = C_{\alpha\alpha}^{1/2}$.

Again let us begin by considering the limiting cases.
In the case of negligible sample covariance
\begin{eqnarray}
F_{\alpha\beta} &=& \sum_i {1 \over \bar  m_i}  \bar m_{i,\alpha} \bar m_{i,\beta}
= \bar{\bf m}^t_{,\alpha} \bar {\bf M}^{-1}
 \bar{\bf m}_{,\beta} \,,
\end{eqnarray}
where $, \equiv \partial/\partial p_\alpha$ \cite{HolHaiMoh01}.
In the case of negligible Poisson errors \cite{TegTayHea97}
\begin{eqnarray}
F_{\alpha\beta} =
 \bar{\bf m}^t_{,\alpha} {\bf S}^{-1}
 \bar{\bf m}_{,\beta} 
  + {1\over 2} {\rm Tr} [{\bf S}^{-1} {\bf S}_{,\alpha} {\bf S}^{-1} {\bf S}_{,\beta}]
\,.
\end{eqnarray}
Notice that even in the $m_i \gg 1$ Gaussian limit, the Poisson Fisher matrix does not carry
a term involving the derivatives of the variance ${\bf M}_{,\alpha}$.  
However the fractional error induced
by including such a term scales as $m_i^{-1}$ and is therefore negligible in this limit.

Given these two limits, we approximate the full Fisher matrix as 
\begin{eqnarray}
F_{\alpha\beta}&=&  \bar{\bf m}^t_{,\alpha} {\bf C}^{-1}
 \bar{\bf m}_{,\beta} 
+ {1\over 2} {\rm Tr} [{\bf C}^{-1} {\bf S}_{,\alpha}
 {\bf C}^{-1} {\bf S}_{,\beta} ]\,,
\end{eqnarray}
where recall ${\bf C}={\bf S}+\bar {\bf M}$.
The two pieces represent the contribution to the information on the parameters from
the mean of the cell counts and their cell-to-cell (co)variance induced by structure in 
the universe.

\section{Self-Calibration}
\label{sec:self}

The two sources of information, the mean counts and their cell-to-cell variance, depend differently
on 
the cluster mass.  By comparing the two one can in principle 
solve for the mass and hence remove the calibration uncertainty in the interpretation of the
number counts.

Given an initial power spectrum,
simulations  can reliably predict the number density of dark matter halos
associated with clusters of a given mass.  
For illustrative purposes, we will employ
the fitting function \cite{Jenetal01}
\begin{equation}
{d \bar n \over d\ln M} = 0.3 {\rho_{m} \over M} {d \ln \sigma^{-1} \over d\ln M}
        \exp[-|\ln \sigma^{-1} + 0.64|^{3.82}]\,,
\label{eqn:massfun}
\end{equation}
where $\sigma^2(M;z)\equiv \sigma^2_{R}(z)$, the density field
variance in a region enclosing $M=4\pi R^3\rho_m/3$ 
at the mean matter density today $\rho_{m}$.  Likewise the bias
of these objects can be described by \cite{MoWhi96,SheTor99}
\begin{equation}
b(M;z) = 1 + {a \delta_c^2/\sigma^2 -1 \over \delta_c} 
         + { 2 p \over \delta_c [ 1 + (a \delta_c^2/\sigma^2)^p]}
\label{eqn:bias}
\end{equation}
with $a=0.75$,  $p= 0.3$,  and $\delta_c=1.69$.  Though these relations must
be replaced by numerical results in a real analysis,  self-calibration will still be possible
so long as the bias and mean number counts scale with the
mass of the objects in a predictable way.  

Consider now a selection that is defined by a threshold in some observable
quantity such flux or temperature.  
Let us define the relationship
between the observable threshold $f_{\rm th}$ and the mass threshold by two parameters
$A$ and $m$ \cite{MajMoh03}
\begin{equation}
{M_{\rm th} \over M_0(z)} = e^{A} (1+z)^m  \left( { f_{\rm th} \over f_0 }\right)^p\,,
\end{equation}
where $f_0$ is an arbitrary normalization parameter, $M_0(z)$ characterizes
an a priori guess for the relation such that deviations can be described by
a power law in $(1+z)$, and $p$ is considered
known (see \cite{Hu03a} for a generalization to arbitrary $p(z)$).  
Then the statistical model of the counts is defined through Eqn.~(\ref{eqn:nz}) and
(\ref{eqn:covariance})
\begin{eqnarray}
\bar n(z) & = & \int_{\ln M_{\rm th}(z)}^\infty d\ln M {d \bar n \over d\ln M}\,, \nonumber\\
b(z)    & = & {1 \over \bar n}  \int_{\ln M_{\rm th}(z)}^\infty d\ln M {d \bar n \over d\ln M} b(M;z)\,.
\end{eqnarray}
In reality the mass-observable relation will contain finite scatter which will blur the threshold; this
scatter must also be modeled in a real analysis.  We use this simple prescription for
illustrative purposes only. 

Fig.~\ref{fig:massfn} illustrates the self-calibration idea.  If only the counts are
considered, changes in the cosmology
are degenerate with those in the threshold.  
However lowering the threshold to compensate a smaller amplitude of
density fluctuations has two effects on the variance of counts that break the
degeneracy: it lowers the variance
due to the decreased underlying linear structure but also makes objects at a fixed
mass rarer and hence more highly biased.

We approximate the results of a joint likelihood analysis of the mass-observable and 
cosmological parameters via the Fisher matrix.
We take the cosmological parameters 
as the normalization of the initial curvature spectrum 
$\delta_\zeta (=5.07\times 10^{-5})$ at $k=0.05$ Mpc$^{-1}$
(see \cite{HuJai03} for its relationship to the more
traditional $\sigma_8$ normalization), its tilt
$n (=1)$, the baryon density
$\Omega_bh^2 (=0.024)$, the dark matter density
$\Omega_m h^2 (=0.14)$,
and the 2 dark energy
parameters of interest: its density
$\Omega_{\rm DE} (=0.73)$ and equation of state $w(=-1)$ which we  assume
 to be constant.
Values in the fiducial cosmology are given in parentheses.  The first 4 parameters
have already been determined at the few to $10\%$ level 
through the CMB \cite{Speetal03} and
we will extrapolate these constraints into the future with priors of
$\sigma(\ln \delta_\zeta)=\sigma(n)=\sigma(\ln\Omega_b h^2)=\sigma(\ln\Omega_m h^2)
=0.01$.

\begin{figure}[tb]
\centerline{\epsfxsize=3.2in\epsffile{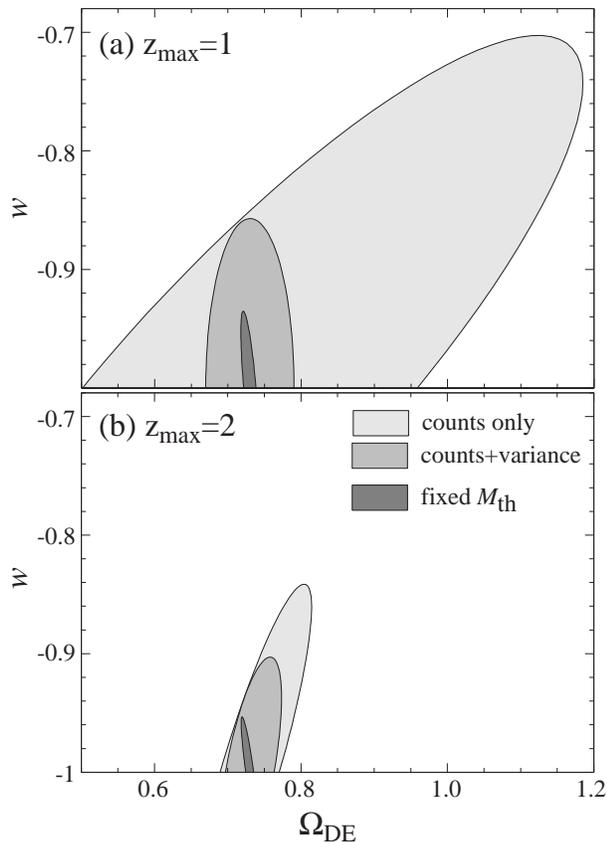}}
\caption{\footnotesize
Projected constraints in the $\Omega_{\rm DE}-w$ dark energy plane (68\% CL) for
the fiducial survey and a maximum redshift of (a) $z_{\rm max}=1$ and (b) $z_{\rm max}=2$.
Unknown power law evolution in the mass-observable relation degrades constraints from
the inner most to outermost ellipses if only the abundance or counts is used.  By adding in
information from the count variance, the survey partially self-calibrates leading to the
middle ellipses.}
\label{fig:wl}
\end{figure}

For illustrative purposes let us take a fiducial cluster survey with specifications
similar to the planned South Pole Telescope (SPT) Survey:
an area of 4000
deg$^{2}$ and a sensitivity corresponding to a constant
$M_{0}$ = $M_{\rm th}|_{\rm fid} = 10^{14.2} h^{-1} M_{\odot}$.
We divide the number counts into bins of redshift $\Delta z=0.1$ and 400 angular
cells of 10 deg$^{2}$ and vary the maximum redshift $z_{\rm max}$ for which photometric
redshifts will be available.  With these large cell sizes, the covariance between neighboring
cells is negligible, considerably simplifying the analysis.

Constraints in the  $\Omega_{\rm DE}-w$ dark energy
plane from counts alone are severely compromised by mass threshold uncertainties 
(see Fig.~\ref{fig:wl}).  This is especially true
for the lower $z_{\rm max}$ since the lever arm is not sufficient to distinguish between
the power law deviations in the mass-observable relation and the dark energy.  Here
the errors in $\Omega_{\rm DE}$  degrade by factor of 37 and $w$ by a factor of 4.6.
Much of the dark energy information is restored by including variance information.  
For $z_{\rm max}=1$, errors improve by a factor of 7.5  to $\sigma(\Omega_{\rm DE})=0.04$ and a factor of 2 to $\sigma(w)=0.09$.

Self-calibration also makes the errors more robust to $z_{\rm max}$, for
$z_{\rm max}=2$,  $\sigma(\Omega_{\rm DE})=0.03$ and  $\sigma(w)=0.06$.
These results are also relatively insensitive to the cell size.  For 4 deg$^{2}$ cells or
$\Delta z= 0.05$  the errors remain nearly the
same.  Likewise, the relative improvement due to self-calibration is relatively insensitive
to the mass threshold assumed.
Finally errors on the actual calibration parameters are $\sigma(A)=\sigma(m)=0.14$ for
$z_{\rm max}=2$. 

\section{Discussion}

Cluster number count surveys contain information not only in the mean counts
 but also in their cell-to-cell variance.  
The latter depends on the clustering of clusters and hence provides
an independent constraint on their mass.  This information automatically comes out of a full
likelihood analysis of the counts and 
provides an opportunity for self-calibration of the survey.

In the idealized case where the scatter in the mass-observable relation is low and known, 
self-calibration via the variance improves dark energy constraints  for power law evolution in the mass threshold by a factor of 2-10
depending mainly on the maximum redshift for which accurate
photometric redshifts can be obtained.  In principle, self-calibration can 
constrain a more arbitrary evolution, e.g. an independent mass-observable relation for
each redshift slice.  In practice, the resulting dark energy constraints  are then too weak to be of interest 
($\sigma(\Omega_{\rm DE})=0.07$,  $\sigma(w)=0.57$ in the fiducial survey).  
Likewise, for an unknown scatter in the mass-observable relation, variance
self-calibration alone is unlikely to suffice.  In these cases it can be supplemented with
selections at various thresholds in the observable \cite{Hu03a} or with
the full angular power spectra of the cell counts (e.g. \cite{HuJai03}) though a full
treatment is beyond the scope of this work.

Since self-calibration involves only information which 
exists in the survey itself, it comes at no additional observational 
cost. It therefore complements potentially more precise but costly
cross-calibration studies.

\smallskip
\noindent{\it Acknowledgments:}  We thank A. Kravtsov and J. Mohr for useful discussions.
This work was supported by NASA NAG5-10840, the DOE, the Packard Foundation and CNPq;
it was carried out  at  the CfCP under NSF PHY-0114422.

\end{document}